%% file: DESY-03-176.tex
\documentclass[zpreprint,zbstpl]{./LaTeX/zeus/zeus_paper}
\usepackage[english]{babel}

\input ./LaTeX/zeus/zeus_def.tex
\input  {DESY-03-176-cit.tex} 
\begin{document}
\include{DESY-03-176-tit}
\include{auth111_out}
\include{DESY-03-176-txt}
\include{DESY-03-176-ref}
\include{DESY-03-176-tab}

\include{DESY-03-176-fig}
%
%
\end{document}

%% file: LaTeX/zeus/zeus_def.tex
\newcommand{\ZcoosysB}{%
The ZEUS coordinate system is a right-handed Cartesian system, with the $Z$
axis pointing in the proton beam direction, referred to as the ``forward
direction'', and the $X$ axis pointing left towards the centre of HERA.
The coordinate origin is at the nominal interaction point.\xspace}

\newcommand{\ZcoosysfnBeta}{\footnote{\ZcoosysB}} 
\newcommand{\Zdetdesc}{%
A detailed description of the ZEUS detector can be found 
elsewhere~\cite{zeus:1993:bluebook}. A brief outline of the 
components that are most relevant for this analysis is given
below.\xspace}
\newcommand{\Zctddesc}[1]{%
Charged particles are tracked in the central tracking detector (CTD)~\citeCTD,
which operates in a magnetic field of $1.43\Tesla$ provided by a thin 
superconducting solenoid. The CTD consists of 72~cylindrical drift chamber 
layers, organized in nine superlayers covering the polar-angle#1 region 
\mbox{$15^\circ<\theta<164^\circ$}. The transverse-momentum resolution for
full-length tracks is $\sigma(p_T)/p_T=0.0058p_T\oplus0.0065\oplus0.0014/p_T$,
with $p_T$ in $\Gev$.}
\newcommand{\Zcaldesc}{%
The high-resolution uranium--scintillator calorimeter (CAL)~\citeCAL consists 
of three parts: the forward (FCAL), the barrel (BCAL) and the rear (RCAL)
calorimeters. Each part is subdivided transversely into towers and
longitudinally into one electromagnetic section (EMC) and either one (in RCAL)
or two (in BCAL and FCAL) hadronic sections (HAC). The smallest subdivision of
the calorimeter is called a cell.  The CAL energy resolutions, as measured under
test-beam conditions, are $\sigma(E)/E=0.18/\sqrt{E}$ for electrons and
$\sigma(E)/E=0.35/\sqrt{E}$ for hadrons, with $E$ in $\Gev$.}

\chardef\usc=95
\chardef\til=126
\catcode`\@=11 
\DeclareRobustCommand\xdotspace{\futurelet\@let@token\@xdotspace}
\def\@xdotspace{%
  \ifx\@let@token.\else
  \ifx\@let@token\bgroup.\else
  \ifx\@let@token\egroup.\else
  \ifx\@let@token\/.\else
  \ifx\@let@token\ .\else
  \ifx\@let@token~.\else
  \ifx\@let@token!.\else
  \ifx\@let@token,.\else
  \ifx\@let@token:.\else
  \ifx\@let@token;.\else
  \ifx\@let@token?.\else
  \ifx\@let@token/.\else
  \ifx\@let@token'.\else
  \ifx\@let@token).\else
  \ifx\@let@token-.\else
  \ifx\@let@token\@xobeysp.\else
  \ifx\@let@token\space.\else
  \ifx\@let@token\@sptoken.\else
   .\space
   \fi\fi\fi\fi\fi\fi\fi\fi\fi\fi\fi\fi\fi\fi\fi\fi\fi\fi}
\catcode`\@=12 

\newcommand{\stru}[2]{%
   \relax\ifmmode\hbox{\vrule height#1 depth#2 width0pt}%
   \else\vrule height#1 depth#2 width0pt\fi}

\newcommand{\Ronum}[1]{\uppercase\expandafter{\romannumeral#1}}
\newcommand{\ronum}[1]{\expandafter{\romannumeral#1}}
\DeclareRobustCommand{\LaTeXZ}{%
  \LaTeX\kern-.05em4\kern-.1em
  {\raisebox{-0.2ex}{$\scriptstyle\text{ZEUS}$}}\xspace}

\DeclareMathAlphabet{\mathbf}{OT1}{cmr}{bx}{sl}
\newcommand{\eVdist}{\kern-0.06667em}

\newcommand{\Gev}{{\text{Ge}\eVdist\text{V\/}}}

\newcommand{\gev}{{\,\text{Ge}\eVdist\text{V\/}}}

\newcommand{\cm}{\,\text{cm}}

\newcommand{\Tesla}{\,\text{T}}

\newcommand{\slashfrac}[2]{%
  \raisebox{0.5ex}{\ensuremath #1}\kern-0.12em/\kern-0.08em
  \raisebox{-.8ex}{\ensuremath #2}}

\newcommand{\sqr}[3]{%
    {\vcenter{\hrule height.#3ex\hbox{\vrule width.#2ex height#1ex
     \kern#1ex\vrule width.#3ex}\hrule height.#2ex}}}

\catcode`\@=11 
\newcommand{\parenbar}{\mathpalette\p@renb@r}
\def\p@renb@r#1#2{\vbox{%
  \ifx#1\scriptscriptstyle \dimen@.7em\dimen@ii.2em\else
  \ifx#1\scriptstyle \dimen@.8em\dimen@ii.25em\else
  \dimen@1em\dimen@ii.4em\fi\fi \offinterlineskip
  \ialign{\hfill##\hfill\cr
    \vbox{\hrule width\dimen@ii}\cr
    \noalign{\vskip-.3ex}%
    \hbox to\dimen@{$\mathchar300\hfil\mathchar301$}\cr
    \noalign{\vskip-.3ex}%
    $#1#2$\cr}}}
\catcode`\@=12 

\newcommand{\IP}{{\rm I$\kern-0.01667em$P}\xspace}

\mathchardef\qsm=63
\mathchardef\pls=43
\mathchardef\mns=512
\mathchardef\plm=518
\mathchardef\eql=61
\mathchardef\smallleft=300
\mathchardef\smallright=301
\mathchardef\les=316
\mathchardef\gre=318
\mathchardef\leq=532
\mathchardef\grq=533

\catcode`\@=11 
\newcounter{pict@width}
\newcounter{pict@height}
\newlength{\pict@scale}
\newcommand{\psfigadd}[4]{%
\setcounter{pict@width}{1*\ratio{#2+\pict@scale/2}{\pict@scale}}
\setcounter{pict@height}{1*\ratio{#3+\pict@scale/2}{\pict@scale}}
\setlength{\unitlength}{\pict@scale}
\hbox to #2{\hspace{-\fill}\begin{picture}(\thepict@width,\thepict@height)
\put(0,0){\psfig{figure=#1,width=#2,height=#3,clip=}}
\SetScale{0.283466457}
\SetWidth{1.763889}
{#4}
\end{picture}}
}
\newcounter{pict@widthfst}
\newcounter{pict@widthscd}
\newcounter{pict@widthtot}
\newcommand{\psfigaddtwo}[7]{%
\setcounter{pict@widthfst}{1*\ratio{#2+\pict@scale/2}{\pict@scale}}
\setcounter{pict@widthscd}{1*\ratio{#2+#4+\pict@scale/2}{\pict@scale}}
\setcounter{pict@widthtot}{1*\ratio{#2+#4+#6+\pict@scale/2}{\pict@scale}}
\setcounter{pict@height}{1*\ratio{#3+\pict@scale/2}{\pict@scale}}
\setlength{\unitlength}{\pict@scale}
\hbox{\hspace{-\fill}\begin{picture}(\thepict@widthtot,\thepict@height)
\put(0,0){\psfig{figure=#1,width=#2,height=#3,clip=}}
\put(\thepict@widthscd,0){\psfig{figure=#5,width=#6,height=#3,clip=}}
\SetScale{0.283466457}
\SetWidth{1.763889}
{#7}
\end{picture}}
}
\newcommand{\psfigror}[4]{%
\setcounter{pict@width}{1*\ratio{#2+\pict@scale/2}{\pict@scale}}
\setcounter{pict@height}{1*\ratio{#3+\pict@scale/2}{\pict@scale}}
\setlength{\unitlength}{\pict@scale}
\hbox{\begin{picture}(\thepict@width,\thepict@height)
\put(0,\thepict@height){\psfig{figure=#1,width=#3,height=#2,clip=,angle=270}}
\SetScale{0.283466457}
\SetWidth{1.763889}
{#4}
\end{picture}}
}
\newcommand{\psfigrol}[4]{%
\setcounter{pict@width}{1*\ratio{#2+\pict@scale/2}{\pict@scale}}
\setcounter{pict@height}{1*\ratio{#3+\pict@scale/2}{\pict@scale}}
\setlength{\unitlength}{\pict@scale}
\hbox{\begin{picture}(\thepict@width,\thepict@height)
\put(0,0){\psfig{figure=#1,width=#3,height=#2,clip=,angle=90}}
\SetScale{0.283466457}
\SetWidth{1.763889}
{#4}
\end{picture}}
}
\catcode`\@=12 
\newlength\listtextwidth

\catcode`\@=11 
\newlength{\@tabfninsert}
\newlength{\@tabfnwidth}
\newcommand{\tabfootnote}[2]{%
  \setlength{\@tabfninsert}{0.8em}
  \setlength{\@tabfnwidth}{\textwidth}
  \addtolength{\@tabfnwidth}{-\@tabfninsert}
  \addtolength{\@tabfnwidth}{-0.4em}
  \noindent\makebox[\@tabfninsert][r]{\footnotesize$^{#1}$\hfil}\hfill%
  \parbox[t]{\@tabfnwidth}{\footnotesize #2\hfill}}
\catcode`\@=12 

%% file: DESY-03-176-cit.tex
\def\citeCTD{{\cite{%
nim:a279:290,*npps:b32:181,*nim:a338:254%
}}\xspace}
\def\citeCAL{{\cite{%
nim:a309:77,*nim:a309:101,*nim:a321:356,*nim:a336:23%
}}\xspace}

%% file: DESY-03-176-tit.tex
\prepnum{DESY 03-176}

\title{
Bose-Einstein correlations in one and two \\ 
dimensions in deep inelastic
scattering}

\author{ZEUS Collaboration}
\date{12 \ December,  2003}

\abstract{
Bose-Einstein correlations in one and two dimensions
have been studied in deep inelastic $ep$ scattering events measured  with the
ZEUS detector at HERA using an integrated luminosity of 121 pb$^{-1}$.
The correlations
are  independent of the virtuality of the exchanged photon, $Q^2$, in the
range $0.1<Q^2<8000 \gev^2$.
There is no significant difference between the correlations in 
the current
and target regions of the Breit
frame for $Q^2>100 \gev^2$.  
The two-dimensional shape of the
particle-production source was investigated, and  
a significant difference between the transverse and
the longitudinal dimensions of the source is observed.
This difference also shows no  $Q^2$ dependence. 
The results
demonstrate that Bose-Einstein interference, 
and hence the size of the particle-production source,
is insensitive to the hard subprocess.
}

\makezeustitle

%% file: auth111_out.tex
\pagenumbering{Roman}                                                                              
                                                   %
\begin{center}                                                                                     
{                      \Large  The ZEUS Collaboration              }                               
\end{center}                                                                                       
  S.~Chekanov,                                                                                     
  M.~Derrick,                                                                                      
  D.~Krakauer,                                                                                     
  J.H.~Loizides$^{   1}$,                                                                          
  S.~Magill,                                                                                       
  S.~Miglioranzi$^{   1}$,                                                                         
  B.~Musgrave,                                                                                     
  J.~Repond,                                                                                       
  R.~Yoshida\\                                                                                     
 {\it Argonne National Laboratory, Argonne, Illinois 60439-4815}, USA~$^{n}$                       
\par \filbreak                                                                                     
  M.C.K.~Mattingly \\                                                                              
 {\it Andrews University, Berrien Springs, Michigan 49104-0380}, USA
% ...
\newpage                                                                                           
$^{\    1}$ also affiliated with University College London, London, UK \\                          
% ...
                                                           %
                                                           %
\newpage   
                                                           %
                                                           %
\begin{tabular}[h]{rp{14cm}}                                                                       
$^{a}$ &  supported by the Natural Sciences and Engineering Research                               
          Council of Canada (NSERC) \\                                                             
$^{b}$ &  supported by the German Federal Ministry for Education and                               
          Research (BMBF), under contract numbers HZ1GUA 2, HZ1GUB 0, HZ1PDA 5, HZ1VFA 5\\         
$^{c}$ &  supported by the MINERVA Gesellschaft f\"ur Forschung GmbH, the                          
          Israel Science Foundation, the U.S.-Israel Binational Science                            
          Foundation and the Benozyio Center                                                       
          for High Energy Physics\\                                                                
$^{d}$ &  supported by the German-Israeli Foundation and the Israel Science                        
          Foundation\\                                                                             
$^{e}$ &  supported by the Italian National Institute for Nuclear Physics (INFN) \\                
$^{f}$ &  supported by the Japanese Ministry of Education, Culture,                                
          Sports, Science and Technology (MEXT) and its grants for                                 
          Scientific Research\\                                                                    
$^{g}$ &  supported by the Korean Ministry of Education and Korea Science                          
          and Engineering Foundation\\                                                             
$^{h}$ &  supported by the Netherlands Foundation for Research on Matter (FOM)\\                   
$^{i}$ &  supported by the Polish State Committee for Scientific Research,                         
          grant no. 620/E-77/SPB/DESY/P-03/DZ 117/2003-2005\\                                      
$^{j}$ &  partially supported by the German Federal Ministry for Education                         
          and Research (BMBF)\\                                                                    
$^{k}$ &  partly supported by the Russian Ministry of Industry, Science                            
          and Technology through its grant for Scientific Research on High                         
          Energy Physics\\                                                                         
$^{l}$ &  supported by the Spanish Ministry of Education and Science                               
          through funds provided by CICYT\\                                                        
$^{m}$ &  supported by the Particle Physics and Astronomy Research Council, UK\\                   
$^{n}$ &  supported by the US Department of Energy\\                                               
$^{o}$ &  supported by the US National Science Foundation\\                                        
$^{p}$ &  supported by the Polish State Committee for Scientific Research,                         
          grant no. 112/E-356/SPUB/DESY/P-03/DZ 116/2003-2005,2 P03B 13922\\                       
$^{q}$ &  supported by the Polish State Committee for Scientific Research,                         
          grant no. 115/E-343/SPUB-M/DESY/P-03/DZ 121/2001-2002, 2 P03B 07022\\                    
\end{tabular}                                                                                      
                                                           %
                                                           %

%% file: DESY-03-176-txt.tex
\pagenumbering{arabic} 
\pagestyle{plain}
\section{Introduction}
\label{sec-int}

The quantum-mechanical wave-function for  a pair of identical bosons 
is  symmetric under particle exchange. 
As a consequence,
interference effects are expected between identical bosons 
emitted close to one another in phase space.
These effects enhance the two-particle density at small
phase-space separations. Such  
Bose-Einstein (BE) correlations for like-charged hadrons
were first observed by Goldhaber et al. 
\cite{prl:1959:181,*pr:1960:300} in $\bar{p}p$ annihilation.

The BE correlations in momentum space
are related to the spatial dimensions of the production source.  
Therefore, studies of the BE effect
may lead to a better understanding of the structure of the source
of the identical bosons.
In deep inelastic scattering (DIS), 
the production volume may depend on the 
virtuality of the exchanged photon, $Q^2$,
since the transverse size of the virtual photon decreases with
increasing $Q^2$.
Under this hypothesis, the BE correlations would depend on $Q^2$.

Alternatively, the size of the region over which BE correlations take
place may be determined only by the soft, or fragmentation, stage of the process.
The BE effect is independent of $Q^2$ if hard scattering and fragmentation 
factorise. For example,      
in the Lund fragmentation model \cite{pl:169B:362,
*apl:b29:1885,*Andersson:1998hs,Andersson:1998xd},
no sensitivity to $Q^2$ is expected,
and the BE correlations between
identical bosons are  a measure of the tension of the colour 
string between partons.  
      
This paper reports on investigations of  the 
BE correlations in one and two dimensions 
in neutral current $ep$ DIS, focusing on  
the dependence of the BE interference
on $Q^2$.  The BE effect in one dimension is  
measured with a higher precision 
than previously in $ep$ collisions \cite{zfp:c75:437} and over a much wider 
kinematic range,  from $Q^2\simeq 0.1$  GeV$^2$ to   
$8000 \gev^2$. 
The correlations are also studied for the first time in DIS in the 
current and the target fragmentation 
regions of the Breit frame  \cite{feynman:1972:photon, *zpf:c2:237},   
which are known to have
rather different fragmentation properties \cite{epj:c11:251}.
The two-dimensional analysis provides sensitivity to 
a possible elongation of the source
expected in the Lund string model \cite{Andersson:1998xd}.

\section{Definition of measured quantities and \\ model predictions}
\label{sec:def}

Bose-Einstein correlations are usually 
parameterised using a Gaussian expression
for the normalised two-particle density \cite{proc:lesip:1984:115}:
\begin{equation}
R (Q_{12}) =\alpha \>  (1+\beta  \>  Q_{12})
\>  (1+ \lambda  \>  e^{- r^2\> Q_{12}^2 } ), 
\label{be1}
\end{equation}
where $Q_{12}\equiv \sqrt{ - (p_1-p_2)^2 } 
= \sqrt{ M^2 - 4 m_{\mathrm{boson}}^2 }$
is the Lorentz-invariant momentum difference between the two identical 
bosons, 
which is related to the invariant mass, $M$, of the two particles  
with four-momenta $p_1$ and $p_2$ and mass $m_{\mathrm{boson}}$.  
In the present analysis, all charged particles are assumed to be pions. 
The parameter   
$\lambda$ is a measure of the degree of 
coherence, i.e. the fraction of pairs of identical particles 
that undergo interference.
For a totally coherent emission of pions, $\lambda=0$, while
for an incoherent source, the symmetrisation of the wave function for 
identical particles  leads to $\lambda=1$.  
The quantity $r$ is the radius of the production volume, 
while the phenomenological parameter $\beta$ is used to  
take into account any long-distance correlations, and
$\alpha$ is a normalisation constant. 
The Gaussian parameterisation in Eq.~(\ref{be1}) is motivated by the assumption that
the emitting sources of identical bosons are described by a spherical
Gaussian density function.   

When the BE correlations are interpreted in terms of the Lund fragmentation
model, the correlation strength is related 
to the string tension. In this case, the correlations 
should have an approximately exponential 
shape \cite{pl:169B:362,*apl:b29:1885,*Andersson:1998hs,Andersson:1998xd}, i.e.  
$\lambda e^{- r^2\> Q_{12}^2}$ 
in Eq.~(\ref{be1}) should be replaced by  $\lambda^{'}e^{- r^{'} \> Q_{12}}$.

To calculate $R(Q_{12})$, the inclusive two-particle 
density, $\rho = 1/N \cdot \mathrm{d}n_{\mathrm{pairs}} / \mathrm{d}Q_{12}$, 
was used,  
where $n_{\mathrm{pairs}}$ is the number of particle pairs
and  $N$ is the number of events.
The densities  were  calculated  for like-charged 
particle combinations ($\rho (++,--)$) 
and for unlike-charged
combinations ($\rho (+-)$), and  the ratio  computed as 
$\xi = \rho (++,--) / \rho (+-)$. This technique
helps to remove correlations due to the topology and global properties
of the  events contributing to $\rho (++,--)$. 
The quantity $\xi$ contains additional 
short-range correlations, due to resonance decays (contributing  
to $\rho (+-)$), which must also be removed.
To reduce such non-BE effects, 
a  Monte Carlo (MC) sample without
the BE effect was used to calculate $\xi^{\mathrm{MC, noBE}}$, and 
non-Bose-Einstein effects were removed by use of the double ratio, 
$R(Q_{12})=\xi^{\mathrm{data}}/\xi^{\mathrm{MC, noBE}}$.  In this case, it
was assumed that    
the BE effect can be  factorised from other types of correlations, and that 
non-BE correlations are well described by the MC models.

It is also possible to extract the BE parameters by considering pairs of
particles from different events as a reference sample (track-mixing
method).  This method 
was not used in the current analysis since it is difficult
to control the systematic effects arising from $Q^2$ differences
of the events in the sample.
Earlier experiments \cite{rpp:66:481,zfp:c75:437}  have found that values of
$r$ extracted with the track-mixing method is systematically smaller than
those obtained with the method used here.
 
Bose-Einstein correlations can be studied in two dimensions. 
For this, the longitudinally co-moving system (LCMS) \cite{proc:rhip:1991:75}
is often used, since in this frame the correlations have  
a convenient interpretation in the Lund-string model.
For $ep$ collisions, the LCMS can be defined
for each pair of particles with momenta ${\bf p_1}$ and ${\bf p_2}$
as the frame in which the sum of the
two momenta, ${\bf p_1} + {\bf p_2}$,
is perpendicular to the $\gamma^*q$  axis, 
as shown in Fig.~\ref{bec_1}.
The three-momentum difference, 
${\bf Q}=({\bf p_2} - {\bf p_1})$, can be decomposed in the LCMS into
transverse, $Q_{T}$, and  longitudinal, $Q_{L}$, components.
The longitudinal direction is aligned with the direction
of motion of the initial parton. Therefore, in the string model, the 
LCMS is the local rest frame of a string.
In this case, the BE effect can be parameterised using the two-dimensional 
function  
\begin{equation}
R (Q_{T}, Q_{L}) =\alpha \> (1+\beta_t \> Q_{T} + \beta_l \> Q_{L})
\> (1+ \lambda  \> e^{- r^2_{T}\> Q_{T}^2 - r^2_{L}\> Q_{L}^2 } ), 
\label{be2}
\end{equation}
where $r_T$ and $r_L$ are the transverse and longitudinal size of
the boson source. The measurements were done using this parameterisation   
and the same procedure as for the one-dimensional study.  

The Lund model predicts that 
the longitudinal size of the production source is 
larger than the transverse one, $r_L>r_T$ 
\cite{Andersson:1998xd}.
However, the usual implementation 
of the BE effect in the Lund MC models does not contain such an elongation.   

The Breit frame \cite{feynman:1972:photon, *zpf:c2:237}, 
as shown in Fig.~\ref{bec_1}, 
allows the separation of 
the radiation of the outgoing
struck quark  from the proton remnant. Therefore, this frame
can be used to test the sensitivity of the BE effect to different
underlying dynamics.
All particles with negative $p_{Z}^{\mathrm{Breit}}$
form the current region.
These  particles are produced by the fragmentation   of
the struck quark, so that this region is
analogous to a
single hemisphere of  an $e^+e^-$ annihilation event,
while the target region is dominated by the softer
fragmentation of the proton remnants.

\section{Experimental setup}

\Zdetdesc
 
\Zctddesc\ZcoosysfnBeta
 
\Zcaldesc

A presampler \cite{nim:a382:419,*magill:bpre} is mounted in front of FCAL, BCAL and
RCAL. It consists of scintillator tiles that detect particles originating from
showers in
the material between the interaction point and the calorimeter. This information
was used to correct the energy of the scattered electron. The position of electrons
scattered close to the electron beam direction is determined by a scintillator
strip detector (SRTD)~\cite{nim:a401:63}. 

The beam-pipe calorimeter (BPC)  \cite{pl:b407:432,*pl:b487:53} was 
installed 294 cm from the interaction point in order 
to enhance the acceptance of the ZEUS detector for low-$Q^2$ events.
It is a tungsten-scintillator sampling calorimeter with the front
face located at $Z=-293.7\cm$, the centre at $Y=0.0$, and the inner
edge of the active area at $X=4.4\cm$, as close as possible to the
rear beam pipe. The energy resolution 
as determined in test-beam measurements with 1--6 GeV 
electrons is ${\sigma_E}/{E}={17\%}/{\sqrt{E}}$, 
with $E$ in GeV.

\section{Data sample}
\label{sec:data}

Two data samples were used for the present analysis. The data for
$Q^2>4 \gev^2$ were taken 
during the 1996-2000 period and correspond to an integrated luminosity of
$121$  pb$^{-1}$. 
The lepton beam energy was $27.6$ GeV and the proton beam
energy was $820$ GeV (1996-1997) or $920$ GeV (1998-2000).
The second sample consists of low-$Q^2$ events taken
with the BPC. 
This sample corresponds to $3.9$ pb$^{-1}$ taken during 1997.

The kinematic variables $Q^2$
and 
Bjorken $x$ were reconstructed using  
the electron method (denoted by  the subscript $e$), which uses
measurements of the energy and angle
of the scattered lepton, 
the double angle (DA) method \cite{proc:hera:1991:23,*h1_da} or 
the Jacquet-Blondel (JB)  method \cite{proc:epfacility:1979:391}.

The scattered-lepton candidate for the region $Q^2>4 \gev^2$ was
identified from the pattern  of
energy deposits in the CAL \cite{nim:a365:508}. 
The following requirements were imposed: 

\begin{itemize}

\item[$\bullet$]
$Q_e^2>4$ GeV$^2$; 

\item[$\bullet$]
$E_{e^{'}}\geq 10$ GeV, where $E_{e^{'}}$ is the
corrected energy of the scattered lepton measured in the CAL;

\item[$\bullet$]
$40\> \leq\> \delta \>\leq\> 60$ GeV, where
$\delta=\sum E_i(1-\cos\theta_i)$,
$E_i$ is the energy of the $i$th calorimeter
cell, $\theta_i$ is its polar angle 
and the sum runs over all cells;

\item[$\bullet$]
$y_{e}\> \leq\> 0.95$ and $y_{\mathrm{JB}}\> \geq\> 0.04$;
 
\item[$\bullet$]
$\mid Z_{\mathrm{vertex}} \mid  < 50$ cm, where
$Z_{\mathrm{vertex}}$ is the vertex position 
determined from the
tracks; 

\item[$\bullet$]
the position of the scattered-lepton candidate in the RCAL
was required to be outside a box of $\pm 14$ cm 
in $X$ and $Y$. 
\end{itemize}

In total, 6.4 million events were selected. 

The low-$Q^2$ events, selected in the region $0.1<Q_e^2<1.0$ GeV$^2$,  
were reconstructed 
by identifying energy deposits in BPC consistent with a
scattered positron with 
an energy of least 7 GeV. The positron position at 
the BPC front face  had to lie within the fiducial area,  
$5.2 <X<9.3\cm$ and $-2.3 <Y<2.8\cm$. 
Other cuts are identical to those used for the 
data sample at $Q^2>4$ GeV$^2$, except for the
$y_\mathrm{JB}$ cut, which was raised to $y_\mathrm{JB}>0.06$.  
The reconstruction of the Breit frame and the $Q^2$ variable
were performed using variables calculated 
with the electron method. In total, about $100000$  events were selected.  

The measurement uses CTD tracks assigned 
to the primary event vertex. 
Tracks were  required to pass through at least three
CTD superlayers and have
transverse momentum $p_{T}^{\mathrm{lab}} > 150$  MeV.
The approximate pseudorapidity range for 
selected tracks is $|\eta | < 1.75$.  
To ensure that tracks were well reconstructed, track pairs 
were required to satisfy 
$Q_{12}>0.05$ GeV.  
In the kinematic
region used for this analysis, pions constitute about $81\%$ of the tracks,
according to MC expectations.

\section{Reconstruction procedure and \\ acceptance correction}
\label{sec:evsim}

The parameters for the BE effect were determined 
from a fit using
either the Gaussian or exponential parameterisation. 
The reference
sample was calculated using unlike-charged pairs, and then the  MC
was used to remove the effect of resonances, 
as discussed in Sect.~\ref{sec:def}.  
The regions affected by imperfections in the MC simulations
of $K^0_S$ and $\rho^0$  decays  were excluded from the fit.

The measured correlation functions, $R(Q_{12})$ and $R(Q_T,Q_L)$,
were corrected for detector effects
using a bin-by-bin procedure.
The detector correction factor 
was calculated from MC events as $\xi^{\mathrm{gen}} / \xi^{\mathrm{det}}$,
where $\xi^{\mathrm{gen}} (\xi^{\mathrm{det}})$
is the generated (reconstructed) two-particle density 
$\rho (++,--)$ divided by $\rho (+-)$.  No cuts were applied 
on the true MC hadrons. 
The corrections are close to unity ($1.0\pm 0.1$), since some
detector effects cancel in the ratios of the two-particle densities.
Only for the lowest $Q_{12}$ region ($0.05<Q_{12}<0.1$ GeV) the
correction is as large 1.3-1.4.

Since the generator-level  MC events do not have the BE effect,
and are used both for the acceptance correction and for the subtraction
of resonances, the final result is equivalent to 
the detector-level measurement in the restricted kinematic 
region defined by the cuts in Sect.~\ref{sec:data}.  
The effect of these cuts, in
particular those on track momentum and angle, depends upon $Q^2$, since $Q^2$ 
determines the available phase space for particle production. The
MC model containing BE correlations was used to estimate the effect of
extrapolation to the full phase space available in each $Q^2$ bin. The  
effect was found to be small 
($+3.5\%$ for the BE radii) and $Q^2$ independent. This model-dependent 
correction was not included 
in the final results.

The MC events were  generated with the 
ARIADNE 4.08 model \cite{cpc:71:15}
interfaced with HERACLES 4.5.2 \cite{cpc:69:155}
using the DJANGOH 1.1 program \cite{spi:www:djangoh11}
to incorporate first-order electroweak corrections.
The generated events were then  passed  through a full
simulation of the detector using GEANT 3.13 \cite{tech:cern-dd-ee-84-1}
and processed with the same
reconstruction program as the data.
The detector-level MC samples  were then
selected in the same way as the data.

To simulate hadronisation, ARIADNE employs the 
PYTHIA 5.7/JETSET 7.4 program \cite{cpc:82:74}, which  
is based on the Lund string model \cite{prep:97:31}. 
The BE effect, which is treated as
a final-state interaction by redistributing hadron momenta according
to a chosen parameterisation, 
is available as an option in the PYTHIA/JETSET 
program. The default  
parameter setting for ARIADNE does not contain the BE effect. 
For systematic checks,  
the BE correlations were included in the simulation
for the acceptance calculations. The BE effect was
parameterised  by  a Gaussian function with the
parameters as determined by H1 \cite{zfp:c75:437}.

As a systematic check, the HERWIG 6.2 \cite{cpc:67:465} model was used
to calculate the acceptance correction.
The hadronisation stage
in HERWIG is described by a cluster
fragmentation model \cite{np:b238:492,*np:b310:461}.
The BE effect is not implemented in the HERWIG model. 

\section{Systematic uncertainties}

The systematic uncertainties of the measured BE correlations
were determined by changing the selection cuts or the analysis procedure and
repeating the extraction of the BE parameters. The following systematic studies
have been carried out, with the resulting  uncertainty for the BE radius for the 
highest-precision measurement (at $Q^2>4$ GeV$^2$)  given
in parentheses:

\vspace{0.2cm}

\begin{itemize}

\item[$\bullet$]
the event-selection cuts on
$y_{e}$, $y_{JB}$, $Z_{\mathrm{vertex}}$, $\delta$, $E_{e'}$ were varied, and 
the scattered-lepton energy  scale
was changed within its uncertainty $\pm 2\%$ ($^{+0.8\%}_{-0.4\%}$).

\item[$\bullet$]
the DA  method was used to reconstruct  
$Q^2$ and the boost vector to the Breit frame ($-0.4\%$); 

\item[$\bullet$]
the minimum transverse momentum for  
tracks was raised by $50$ MeV ($-3\%$);

\item[$\bullet$]
tracks were required to be within the pseudorapidity 
range $\mid \eta \mid < 1.5$, in
addition to the requirement of three CTD superlayers ($+0.1\%$);

\item[$\bullet$]
the fit was performed 
excluding one data point from each side of the region
affected by $\rho$ decays. In addition, 
the lowest $Q_{12}$ bin,  which is not well described by the Gaussian parameterisation 
was excluded from the fit ($^{+0.3\%}_{-0.4\%}$).

\end{itemize}

\vspace{0.2cm}

The overall systematic uncertainty was  determined by adding
the  above  uncertainties in quadrature. 

As an additional systematic check,  
the HERWIG model was used
to calculate the acceptance correction.
Change in the reconstructed BE radius
was $+9\%$. However,  
HERWIG does not reproduce
the measured $\rho(+-)$ density, so 
the extraction of the BE effect
is less reliable; therefore, this check was not included
in the final systematic uncertainties.

Identically charged particles are subject to Coulomb repulsion,
which is not simulated by MC models. 
The Bose-Einstein correlation function 
was corrected in the data using the Gamow factor \cite{pr:c20:2267}.      
After correcting for the Coulomb effect, 
the size of the BE radius and $\lambda$ slightly increased ($+4\%$ for $r$). 
It was found that this correction does not depend on $Q^2$, therefore,  
it was not included in the final results.

\section{Results and discussion}

\subsection{One-dimensional study}

Figure~\ref{bec_2} shows the measured $R(Q_{12})$ for $Q^2>4 \gev^2$  
together with the Gaussian fit of Eq.~(\ref{be1}) and
the exponential fit.  
The observed distortions for $Q_{12}>0.9$ GeV are caused
by the decay products of resonances which are not well described by
the MC simulation.    
Both parameterisations give fits of similar quality.
The extracted parameters for the  Gaussian and exponential fits  
are given in Table~\ref{tab-1D}. 

The BE parameters extracted using the Gaussian parameterisation 
are shown in Fig.~\ref{bec_3} as functions of $Q^2$. 
Within the statistical and systematic uncertainties, the data indicate 
no variations with the virtuality of the exchanged photon
in the range $0.1< Q^2 < 8000 \gev^2$ when either the
Gaussian or exponential parameterisation is used. 
This measurement is consistent with an earlier H1 
measurement \cite{zfp:c75:437} for  $6<Q^2<100 \gev^2$ using the wrong-charge
background subtraction.  

Figure~\ref{bec_3} also shows the comparisons between the
current and the target regions of the Breit frame. The BE effect
in the current region was extracted for $Q^2>100 \gev^2$,
where the charged multiplicity is high  enough 
for a reliable measurement. 
The result indicates  that 
there is no significant difference
between the BE effects in the current and the target regions
of the Breit frame.
Note that the  data shown in Fig.~\ref{bec_3} for the total phase space 
are dominated by the target
region.

\subsection{Two-dimensional study}

The BE correlations 
can be studied in more than one dimension after decomposing
the momentum difference into its transverse and longitudinal
components. 
The BE parameters, $r_L$, $r_L$ and $\lambda$  
are  shown in Figure~\ref{bec_4}. Table~\ref{tab-2D} also 
gives the ratio $r_T/r_L$.  
The result shows  that the pion-emitting region, as observed in the LCMS,
is elongated with $r_L$ being larger than $r_T$.

Figure~\ref{bec_4} and Table~\ref{tab-2D} also show 
extracted BE parameters as a function of $Q^2$ 
for the two-dimensional measurements. There is again 
no $Q^2$ dependence of the BE radii.

\subsection{Comparisons with other experiments}

One-dimensional BE correlations have been measured in a number of experiments 
using the wrong-charge
background subtraction   
\cite{zfp:c32:1, pl:b286:201,*zfp:c54:75,*zpf:c72:389,*pl:b524:55,*pl:b559:131, epj:c8:559,*pl:b493:233,*pl:b547:139, zp:c54:21ss,*zp:c59:195ss, zfp:c36:517,*pl:b226:410,*zfp:c43:341,*zfp:c54:21,*zp:c59:195, pl:b458:517,*pl:b471:460,*epj:c16:423}. 
For DIS, the present result agrees well
with the $\mu p$ data at $Q^2>4 \gev^2$ measured by the 
EMC Collaboration \cite{zfp:c32:1}, as well as with 
the data for $6<Q^2<100$ GeV$^2$ measured by H1 \cite{zfp:c75:437}, 
as discussed before.

The data  also agree with the LEP1 average 
$r=0.78\pm 0.01(stat.)\pm 0.16(syst.)$ \cite{pl:b452:159} calculated 
by combining results for the track-mixing and wrong-charge
background-subtraction methods,     
as well as with LEP2 measurements \cite{epj:c8:559,*pl:b493:233,*pl:b547:139}. 
The present results also agree
with earlier lower-energy $e^+e^-$ annihilation experiments 
(see a review \cite{ijmp:a4:2861}).
Since the BE effect measured in this paper is dominated by the
target region, this suggests that the BE effect is not
sensitive to the underlying hard processes.  

Comparisons with  $\pi^+p$ and $pp$ data
\cite{zfp:c36:517,*pl:b226:410,*zfp:c43:341,*zfp:c54:21,*zp:c59:195, zp:c54:21ss,*zp:c59:195ss} indicate that
the BE radii  may be  larger for these two processes. 
However, since the observed BE interference significantly depends
on the experimental procedure used to extract the effect, it is difficult
to assess quantitative differences between the BE correlations observed in
DIS and hadron-hadron collisions.

The measured BE effect
reported here disagrees with that in relativistic heavy-ion collisions,
which are characterised by
significantly larger BE radii, which
depend on the atomic number, $A$,
of the projectile as $r\simeq 0.7A^{1/3}$ fm \cite{ijmp:a4:2861}.

The LEP experiments have recently reported
an elongation of the pion source in $e^+e^-$ annihilation events
\cite{pl:b458:517,*pl:b471:460,*epj:c16:423}.
The present ratio for the two-dimensional BE effect in $ep$ collisions is
consistent with these  measurements.  

\section{Conclusions}

One- and two-dimensional Bose-Einstein correlations
have  been studied in deep inelastic $ep$ scattering. 
The effect was measured
as a function of  the photon virtuality, $Q^2$, 
in the range from $0.1$ to $8000 \gev^2$. 
The results indicate that the source of identical
particles has an elongated shape,
consistent with the expectations of the Lund model.  

The Bose-Einstein  effect in one and two dimensions 
does not depend on the virtuality of the 
exchanged photon.  
The elongation of the pion source is also independent
of $Q^2$.   
In addition, the Bose-Einstein  correlations in  
the current and target regions of the Breit frame 
are similar, even though there is a significant difference in the underlying
physics in these two regions.

These high-precision results, obtained over a wide kinematic range 
in a single experiment, 
demonstrate that Bose-Einstein interference
in $ep$ collisions, and hence the size of the particle-production source,
is insensitive to the hard subprocess.

\section*{Acknowledgements}
\vspace{0.3cm}
We thank the DESY Directorate for their strong support and encouragement.
The remarkable achievements of the HERA machine group were essential for
the successful completion of this work and are greatly appreciated. We
are grateful for the support of the DESY computing and network services.
The design, construction and installation of the ZEUS detector have been
made possible owing to the ingenuity and effort of many people from DESY
and home institutes who are not listed as authors.

\vfill\eject 

%% file: DESY-03-176-ref.tex
{
\def\bibname{\Large\bf References}
\def\refname{\Large\bf References}
\pagestyle{plain}
\ifzeusbst
  \bibliographystyle{./BiBTeX/bst/l4z_default}
\fi
\ifzdrftbst
  \bibliographystyle{./BiBTeX/bst/l4z_draft}
\fi
\ifzbstepj
  \bibliographystyle{./BiBTeX/bst/l4z_epj}
\fi
\ifzbstnp
  \bibliographystyle{./BiBTeX/bst/l4z_np}
\fi
\ifzbstpl
  \bibliographystyle{./BiBTeX/bst/l4z_pl}
\fi
{\raggedright
\bibliography{./BiBTeX/user/syn.bib,%
              ./BiBTeX/bib/l4z_articles.bib,%
              ./BiBTeX/bib/l4z_books.bib,%
              ./BiBTeX/bib/l4z_conferences.bib,%
              ./BiBTeX/bib/l4z_h1.bib,%
              ./BiBTeX/bib/l4z_misc.bib,%
              ./BiBTeX/bib/l4z_old.bib,%
              ./BiBTeX/bib/l4z_preprints.bib,%
              ./BiBTeX/bib/l4z_replaced.bib,%
              ./BiBTeX/bib/l4z_temporary.bib,%
              ./BiBTeX/bib/l4z_zeus.bib}}
}
\vfill\eject

%% file: DESY-03-176-tab.tex

\begin{table}[p]
\begin{center}

\begin{sideways}
\input{DESY-03-176-tab1.tex}

\end{sideways} 
 
\caption{
The values of the one-dimensional BE parameters for different $Q^2$ ranges.
The Gaussian and exponential parameterisations were used 
to extract the parameters.
The statistical  and systematic uncertainties are indicated.
}
  \label{tab-1D}
\end{center}

\end{table}

\begin{table}[p]
\begin{center}

\input{DESY-03-176-tab2.tex}
\caption{
The values of the two-dimensional BE parameters for different $Q^2$ ranges,
as well as the ratio $r_T/r_L$.  
The Gaussian parameterisation was used to extract the parameters.
The statistical  and systematic uncertainties are indicated.
 }
\label{tab-2D}
\end{center}
\end{table}

%% file: DESY-03-176-tab1.tex
%
\begin{tabular}{|c|c|c|c|c|}
\hline  
Q$^2$ (GeV$^2$)  &
$\lambda $ & r (fm) & $\lambda^{'}$ & r$^{'}$ (fm) \\
\hline  
\hline  
 4 - 8000 & 0.475 $\pm$ $0.007^{+0.011}_{-0.003}$  & 0.666 $\pm$ $0.009^{+0.022}_{-0.036}$ 
       & 0.913 $\pm$ $0.015^{+0.099}_{-0.005}$ & 0.928 $\pm$ $0.023^{+0.005}_{-0.094}$\\
 
\hline  
 100 - 8000 & 0.431 $\pm$ $0.015^{+0.014}_{-0.013}$ & 0.646 $\pm$ $0.021^{+0.004}_{-0.029}$ 
        &0.815 $\pm$ $0.037^{+0.110}_{-0.014}$ & 0.859 $\pm$ $0.059^{+0.012}_{-0.113}$\\
\hline
\hline  

0.1 - 1 & 0.464 $\pm$ $0.027^{+0.020}_{-0.044}$ & 0.602 $\pm$ $0.036^{+0.020}_{-0.051}$
        & 0.929 $\pm$ $0.069^{+0.076}_{-0.132}$ & 0.785 $\pm$ $0.071^{+0.119}_{-0.075}$\\
\hline  
4 - 8   & 0.468 $\pm$ $0.020^{+0.009}_{-0.006}$ & 0.685 $\pm$ $0.028^{+0.004}_{-0.054}$
        & 0.892 $\pm$ $0.043^{+0.117}_{-0.008}$ & 0.954 $\pm$ $0.069^{+0.015}_{-0.168}$\\
\hline  
8 - 16  & 0.472 $\pm$ $0.016^{+0.029}_{-0.001}$ & 0.620 $\pm$ $0.018^{+0.031}_{-0.038 }$ 
        & 0.911 $\pm$ $0.041^{+0.163}_{-0.004}$ & 0.857 $\pm$ $0.054^{+0.040}_{-0.089}$\\
\hline  
16 - 32  & 0.473 $\pm$ $0.017^{+0.017}_{-0.009}$ & 0.629 $\pm$ $0.022^{+0.007}_{-0.035}$
         & 0.926 $\pm$ $0.052^{+0.174}_{-0.019}$ & 0.829 $\pm$ $0.066^{+0.014}_{-0.113}$\\
\hline  
32 - 64  & 0.496 $\pm$ $0.018^{+0.020}_{-0.013}$ & 0.679 $\pm$ $0.022^{+0.022}_{-0.032}$ 
         & 0.941 $\pm$ $0.042^{+0.100}_{-0.018}$ & 0.910 $\pm$ $0.060^{+0.042}_{-0.076}$\\
\hline  
64 - 128  & 0.445 $\pm$ $0.017^{+0.019}_{-0.003}$ & 0.665 $\pm$ $0.024^{+0.006}_{-0.049}$
          & 0.843 $\pm$ $0.038^{+0.132}_{-0.007}$ & 0.901 $\pm$ $0.063^{+0.006}_{-0.114}$\\
\hline  
128 - 400  & 0.431 $\pm$ $0.021^{+0.010}_{-0.011}$ & 0.649 $\pm$ $0.030^{+0.005}_{-0.036}$ 
           & 0.821 $\pm$ $0.050^{+0.087}_{-0.013}$ & 0.879 $\pm$ $0.081^{+0.004}_{-0.129}$\\
\hline  
400 - 1200  & 0.454 $\pm$ $0.059^{+0.005}_{-0.024}$ & 0.657 $\pm$ $0.080^{+0.016}_{-0.018}$
            & 0.852 $\pm$ $0.139^{+0.081}_{-0.019}$ & 0.889 $\pm$ $0.216^{+0.027}_{-0.066}$ \\
\hline  
1200 - 8000  & 0.446 $\pm$ $0.120^{+0.063}_{-0.086}$ & 0.837 $\pm$ $0.164^{+0.117}_{-0.073}$
             & 0.841 $\pm$ $0.234^{+0.132}_{-0.173}$ & 1.227 $\pm$ $0.389^{+0.237}_{-0.169}$\\
\hline  
\hline
\end{tabular}
%

%% file: DESY-03-176-tab2.tex
\begin{tabular}{|c|c|c|c|c|}
\hline  
Q$^2$ (GeV$^2$)  &
$\lambda $ &  r$_L$  (fm) &  r$_{T} $ (fm)  &  r$_{T}$/r$_{L} $ \\
\hline  
\hline  
 4 - 8000 & 0.44 $\pm$ $0.01^{+0.01}_{-0.03}$  & 0.95 $\pm$ $0.03^{+0.03}_{-0.08}$ 
       & 0.69 $\pm$ $0.01^{+0.01}_{-0.06}$ & 0.72 $\pm$ $0.03^{+0.04}_{-0.03}$\\
 
\hline  
 100 - 8000 & 0.32 $\pm$ $0.03^{+0.02}_{-0.01}$ & 0.88 $\pm$ $0.08^{+0.03}_{-0.06}$ 
        &0.62 $\pm$ $0.04^{+0.05}_{-0.01}$ & 0.70 $\pm$ $0.08^{+0.06}_{-0.01}$\\
\hline
\hline  

0.1 - 1 & 0.41 $\pm$ $0.05^{+0.08}_{-0.00}$ & 0.82 $\pm$ $0.09^{+0.03}_{-0.02}$
        & 0.74 $\pm$ $0.08^{+0.01}_{-0.13}$ & 0.91 $\pm$ $0.14^{+0.03}_{-0.18}$\\
\hline  
4 - 16   & 0.46 $\pm$ $0.02^{+0.06}_{-0.01}$ & 0.84 $\pm$ $0.04^{+0.04}_{-0.03}$
        & 0.69 $\pm$ $0.02^{+0.04}_{-0.02}$ & 0.83 $\pm$ $0.05^{+0.03}_{-0.00}$\\
\hline  
16 - 64  & 0.39 $\pm$ $0.02^{+0.03}_{-0.05}$ & 1.03 $\pm$ $0.07^{+0.20}_{-0.11}$ 
        & 0.66 $\pm$ $0.03^{+0.02}_{-0.02}$ & 0.64 $\pm$ $0.05^{+0.07}_{-0.10}$\\
\hline  
64 - 400  & 0.34 $\pm$ $0.02^{+0.02}_{-0.05}$ & 0.85 $\pm$ $0.07^{+0.21}_{-0.05}$
         & 0.62 $\pm$ $0.03^{+0.03}_{-0.00}$ & 0.73 $\pm$ $0.07^{+0.06}_{-0.16}$\\
\hline  
400 - 8000  & 0.42 $\pm$ $0.10^{+0.06}_{-0.01}$ & 1.08 $\pm$ $0.27^{+0.12}_{-0.00}$
          & 0.67 $\pm$ $0.11^{+0.11}_{-0.03}$ & 0.62 $\pm$ $0.18^{+0.07}_{-0.05}$\\

\hline  
\hline
\end{tabular}
\\

%% file: DESY-03-176-fig.tex

\begin{figure}
\begin{center}
\includegraphics[height=4.5cm]{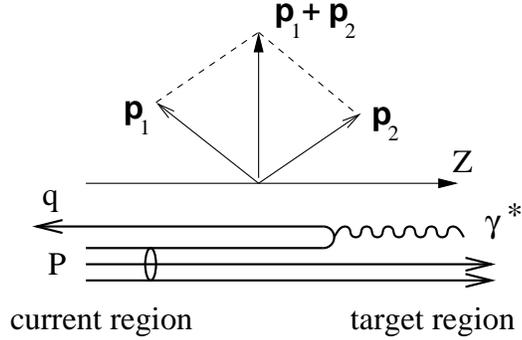}%
\caption{
The longitudinally co-moving system  for a pair of particles in DIS. This system 
is defined as the   
frame of reference in which the sum of the two-particle momenta,
${\bf p}_1$ and ${\bf p}_2$, is perpendicular to the $\gamma^*q$
axis, which coincides with the $Z$ axis of the Breit frame.      
}
\label{bec_1}
\end{center}
\end{figure}

\begin{figure}
\begin{center}
\includegraphics[height=12.0cm]{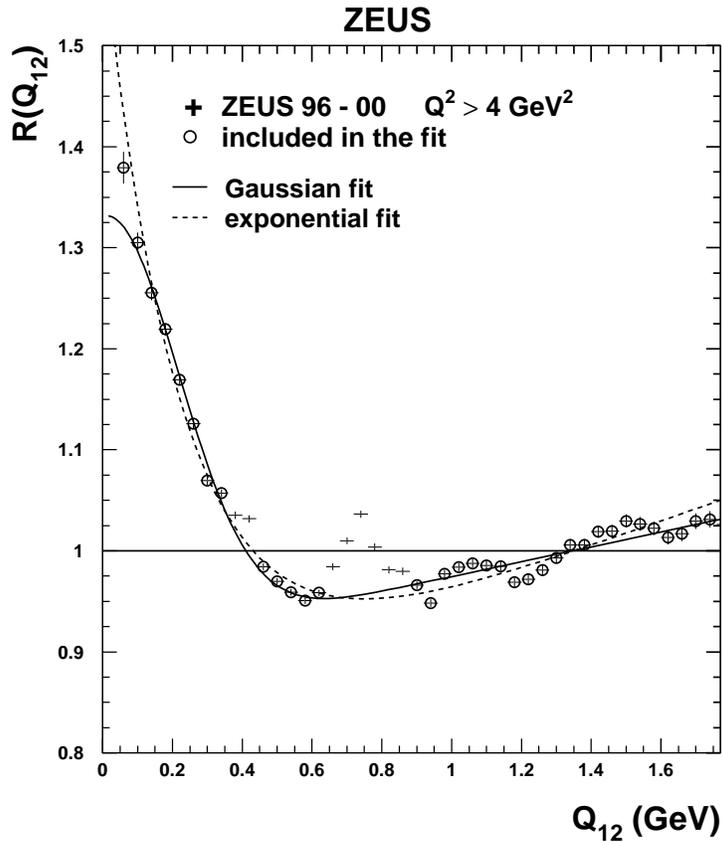}%
\caption{
The measured Bose-Einstein correlation function,  $R(Q_{12})$, together
with the Gaussian and the exponential fits. 
The error bars show the statistical uncertainties.
The data points included in the fit are marked with  the circles.
} 
\label{bec_2}
\end{center}
\end{figure}


\newpage
\begin{figure}
\begin{center}
\includegraphics[height=16.0cm]{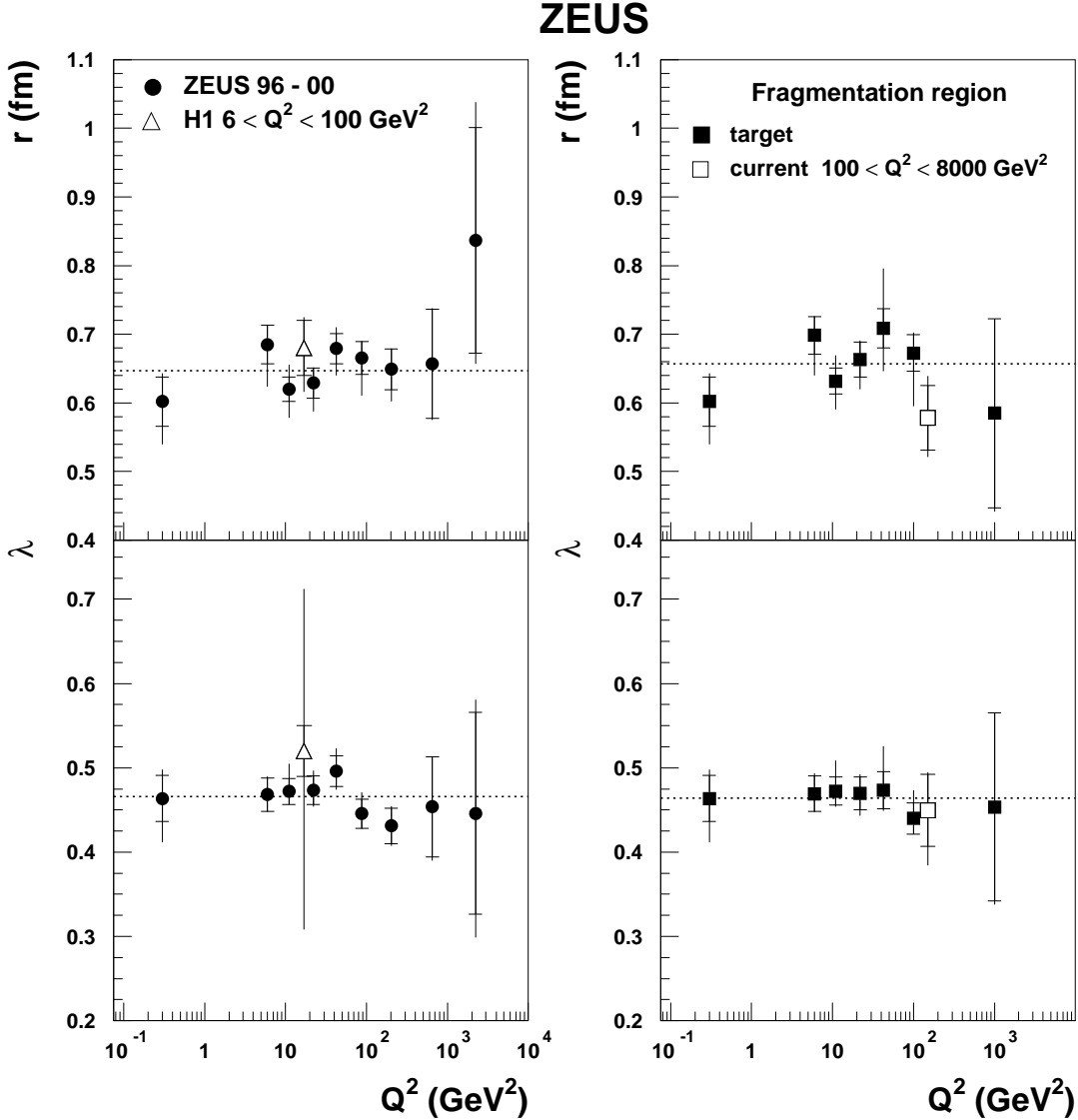}%
\caption{
The extracted 
radius, $r$, and the incoherence parameter, $\lambda$, 
as functions of $Q^2$ for the total measured phase space (left figures) and
for the target region and the current region of the Breit frame (right figures). 
The H1 data point
is shown for the mean value of the measured $Q^2$ range, $6<Q^2<100 \gev^2$.
The BE effect is shown for the current fragmentation region for
$100<Q^2<8000 \gev^2$.
The inner error bars
are statistical uncertainties; the outer are statistical and systematic
uncertainties  added in quadrature.
The dotted lines show the average values, with the $\chi^2/ndf\simeq 0.5$ for
both $r$ and $\lambda$.  
}
\label{bec_3}
\end{center}
\end{figure}

\newpage
\begin{figure}
\begin{center}
\includegraphics[height=15.0cm]{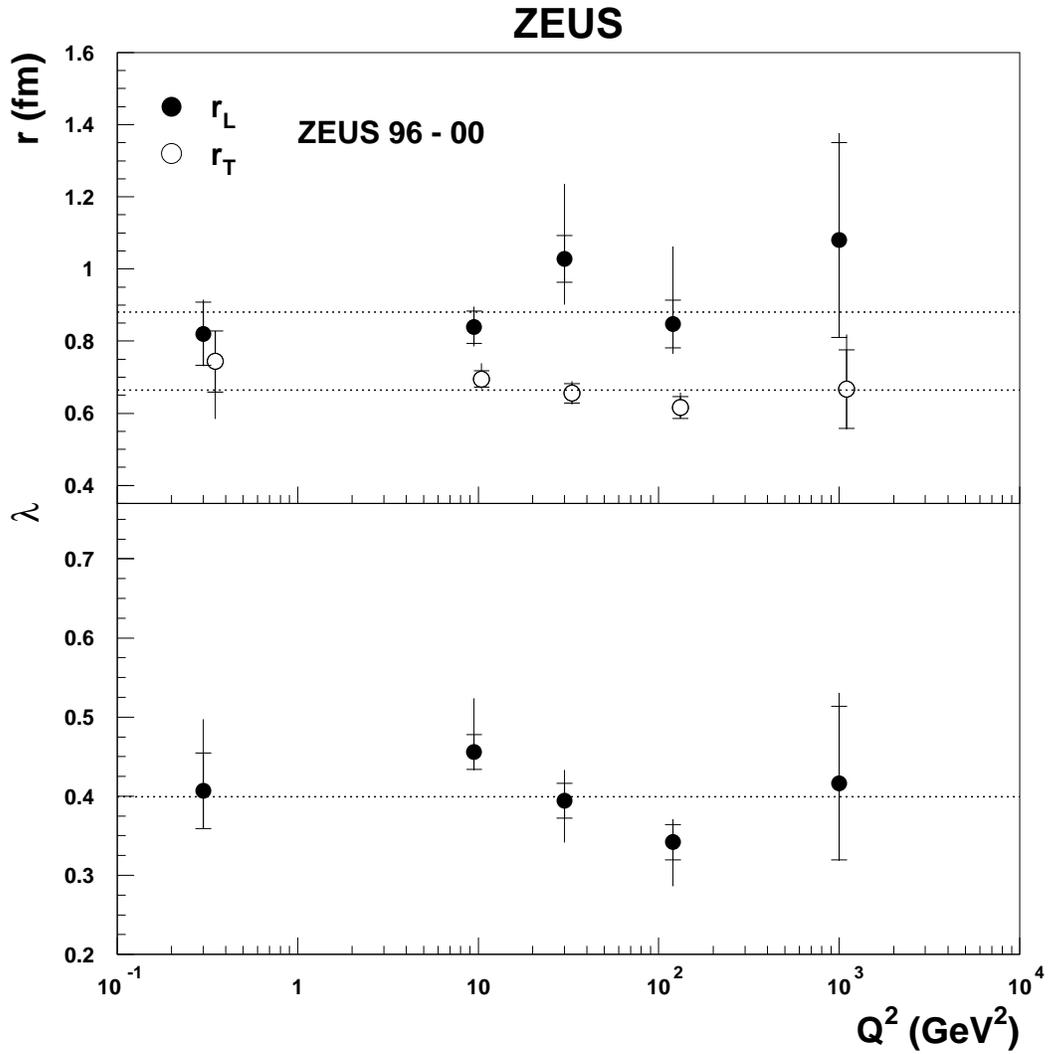}%
\caption{
The extracted radii, $r_T$, $r_L$,  and the
incoherence parameter $\lambda$ as functions of $Q^2$
for the two-dimensional correlation function $R(Q_T, Q_L)$.
The inner error bars
are statistical uncertainties; the outer are statistical and systematic
uncertainties  added in quadrature.
The dotted lines show the average values.
}
\label{bec_4}
\end{center}
\end{figure}